\documentclass{ws-rv9x6}
\usepackage{subfigure}     
\usepackage[square]{ws-rv-van}     
\makeindex
\begin{document}

\chapter[Pairing beyond mean field]{Pairing in finite systems: beyond the HFB theory}\label{ra_ch1}

\author[Robledo and Bertsch ]{L.M. Robledo$^{(1)}$ and G.F. Bertsch$^{(2)}$}

\address{$^{(1)}$Dep. Fisica Te\'orica, Modulo 15  Universidad Aut\'onoma de Madrid, E-28049 Madrid, Spain \\
luis.robledo@uam.es\\
$^{(2)}$Institute for Nuclear Theory and Dept. of Physics, Box 351560, University 
of Washington, Seattle, Washington 98915, USA\\bertsch@uw.edu}

\begin{abstract}
The Hartree-Fock-Bogoliubov approximation is very useful for treating
both long- and short-range correlations in
finite quantum fermion systems, but it must be extended in order
to describe detailed spectroscopic properties.  One
problem is the symmetry-breaking character of the HFB approximation.
We present a general and systematic way to restore symmetries and 
to extend the configuration space
using pfaffian formulas for the many-body matrix elements. The advantage of those
formulas is that the sign of the matrix elements is unambiguously determined.
It is also helpful to extend the space of configurations by constraining the HFB 
solutions in some way.  A powerful method for finding these constrained 
solutions is the gradient method, based on the generalized Thouless transformation.
The gradient method also preserves the number parity of the
Bogoliubov transformation, which facilitates the application
of the theory to systems with odd particle number. 
\end{abstract}
\body

\section{Introduction}\label{ra_sec1}

Soon after the seminal paper describing the microscopic theory of superconductivity by
Bardeen- Cooper- Schrieffer (BCS) \cite{PhysRev.108.1175},  Bohr et al.
\cite{PhysRev.110.936} found an analogy between
the excitation spectra of nuclei and those of the 
superconducting metallic state and pointed out the role of pairing
correlations in the low excitation spectrum of atomic nuclei. 
As self-bound fermionic
systems, nuclei are unique in requiring for their theoretical
description the inclusion of both long- and short-range correlations.  
The longest range correlations may be treated
in the Hartree-Fock (HF) approximation with a suitable effective Hamiltonian.  
In the simplest theory that includes pairing, the
pairing correlations are introduced through the BCS approximation
defining the pair amplitudes from the time-reversed orbital
wave functions of the HF theory.  However, in many situations
the HF/BCS wave functions are not the variational minima in the complete
space of wave functions defined by the Bogoliubov transformations.  
For this reason contemporary
calculations of nuclear structure based on the mean-field approximation
(see \cite{RevModPhys.75.121,Erler.11,Niksic.11} for recent reviews)
largely follow the Hartree-Fock-Bogoliubov (HFB) formulation of theory; see Refs 
\citep{RS80,BR86} for details in the nuclear physics context.

The atomic nucleus is a mesoscopic system where the broken symmetry 
implied by the BCS or HFB wave functions is just an artifact
of the mean field approximation. An improved description of physical properties 
requires techniques beyond the mean field, like particle number symmetry restoration or fluctuations in the BCS 
order parameter. Those techniques were developed in the 1960's 
\cite{Bayman.60,PhysRev.135.B22,Onishi.66,Balian.69} 
and applied to a variety of situations in nuclear physics \cite{RevModPhys.75.121,Erler.11,Niksic.11}.
Recently, they have been exported to several branches of physics \cite{Yannouleas.07} and quantum chemistry \cite{Scuseria.12}. 
Other approaches based on the Random Phase approximation and derivatives are also popular (see Y.R. Shimizu contribution to
this Volume and Ref \cite{BrBr.05} )
However, technical difficulties still remain
in its practical implementation,  especially  in systems where time 
reversal symmetry is broken. One of the difficulties is evaluating  the sign of matrix elements between two general HFB wave functions.
The sign is relevant because it determines the interference pattern of those
linear combinations of mean field wave functions typical of theories for symmetry restoration and/or configuration mixing.
The proof that the sign of the matrix elements is well defined was given in the past \cite{Neergard.83} but
a general and robust  methodology to determine it in practice was not
available until a new method based on pfaffians was introduced \cite{PhysRevC.79.021302}. 
The generalization to systems with an odd number of particles (to be
denoted odd-A systems) has been given recently  \cite{PhysRevLett.108.042505} 
and our methodology will be discussed below.

The HFB theory defines a minimization problem that raises the practical question
of finding the minimum of an energy function that depends on
a large number of variables.  Traditionally 
the equation for the gradient, ie. the derivative of the energy function with
respect to all the variables, is set equal to zero and the resulting
HFB equations are solved iteratively.
However, it has been long known  that there can be severe 
difficulties with this approach, as may be seen in Fig 5.3 of the textbook
by Ring and Schuck \cite{RS80}). 
The approach using the gradient directly is more stable, and we have taken
this path in our group at Madrid  to develop efficient 
codes based on a second-order treatment of the gradient.
One situation where the gradient method has obvious advantages is 
in treating systems with an odd number of particles, discussed in 
Section \ref{Sec2} below.  It is also much easier to treat a large 
number of constraints in the gradient method. This will facilitate 
the extensions of the HFB theory discussed in Section \ref{Sec1} below.

\section{Sign of HFB overlaps with the pfaffian technique}\label{Sec1}

The problem of calculating the overlap of two HFB wave functions was first 
considered in the 1960's \cite{Onishi.66}
in the context of symmetry restoration. The formula derived there involves the square root of the
determinant of a matrix built with the Bogoliubov amplitudes $U$ and $V$ of the HFB states involved.
The presence of the square root implies that the sign is undefined. However, if 
time reversal is preserved, Kramers degeneracy implies that the determinant in the overlap
formula is the square of a number and its sign is usually assigned to the overlap (without proof).
For general HFB states it can be proven \cite{Neergard.83} that the eigenvalues of 
the matrix in the argument of the determinant are doubly degenerate implying that the determinant
is again the square of a number.

\begin{figure}\label{Fig1}
\centerline{\psfig{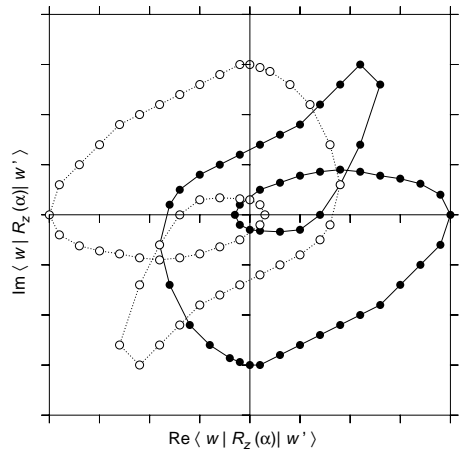}}
\caption{Sketch of the real and imaginary parts of a
typical overlap of the form $\langle w | \hat{R}_z (\alpha) | w ' \rangle$. 
Filled circles are the values of
the overlap; open circles the same but with the opposite sign.}
\label{ra_fig1}
\end{figure}
To illustrate the sign problem we present in Fig \ref{Fig1} a sketch of the real and imaginary parts of a
typical overlap of the form $\langle w | \hat{R}_z (\alpha) | w ' \rangle$ where the angle $\alpha$
varies between 0 and $2\pi$. Realistic examples are presented and discussed, for instance, in Ref 
\citep{PhysRevC.85.034325}. In our sketch plot, two sets of points are depicted.
The filled circles represent the 
overlaps obtained on a discrete mesh  of $\alpha$ values. The open circles are the same 
overlaps but with opposite sign. The lines joining the points are plotted to guide the eye. The
overlaps are used typically in integrals in $\alpha$ (see \citep{PhysRevC.78.024309} for examples).
From the plot it becomes clear that if the procedure to identify the sign is not robust (usually
arguments based on continuity of the overlap as a function of $\alpha$ are used) one can easily
jump onto the wrong curve when the modulus of the overlap is small. At first sight
it could be argued that the error in the integral is going to be small as the jump takes place
in the region of small overlap moduli but continuing in the wrong curve leads 
to large values of the overlaps with the wrong sign.

An unambiguous evaluation of the sign of the overlap between two HFB wave functions
was first achieved in Ref \citep{PhysRevC.79.021302}. That expression for the 
overlap was derived by  the coherent fermion state technique, resulting in a 
pfaffian of a matrix related to the Bogoliubov transformation matrices. While
this solves the problem for fully paired HFB wave functions, the matrix expression
can become singular in the HF limit. Other pfaffian expressions addressing 
this and other problems related with the use of different finite bases for
different states were subsequently found \cite{PhysRevC.84.014307}. 
The limitation in these approaches is that only fully paired
HFB wave functions are allowed and the method is restricted to 
systems with even number parity. 
Recently, a method that uses the expression of the standard Wick theorem for mean values
of fermion operators in the vacuum in terms of a pfaffian has permited the
extension of the previous result to odd-A systems \cite{PhysRevLett.108.042505}. 
Other treatments of  odd-A systems \citep{Oi2012305,PhysRevC.85.034325} require 
the Generalized Wick Theorem (GWT) \cite{Balian.69} and lead to more elaborated expressions.

The results obtained in \cite{PhysRevLett.108.042505} are based on a result
for the expectation values of fermion operators in the vacuum.  The method
may be understood more easily with an example. If ${\beta_i}$  are fermion creation or annihilation operators
satisfying the standard commutation relations, the standard Wick theorem says that the
following mean value with respect to  the vacuum
$$
\langle - | \beta_{1} \beta_{2}  {\beta}_{3}
{\beta}_{4} | - \rangle= r_{12} r_{34} - r_{13} r_{24} + r_{14}r_{23}
$$
is given in terms of the contractions $r_{ij}=\langle - |\beta_i \beta_j | - \rangle$. 
On the other hand, the pfaffian of a general $4\times 4$ (skew-symmetric) matrix 
is given by \footnote{See \cite{PhysRevC.79.021302}  for basic results and bibliography concerning pfaffians
and \cite{Gonzalez-Ballestero2010} for numerical and symbolic techniques.}
$$
 \textrm{pf} \left( \begin{array}{cccc}   0    & r_{12} & r_{13} & r_{14} \\ 
                             -r_{12} &    0   & r_{23} & r_{24} \\
							 -r_{13} &-r_{23} &   0    & r_{34} \\
							 -r_{14} &-r_{24} & -r_{34}&   0
							 \end{array} \right) 
= r_{12}r_{34}-r_{13}r_{24}+r_{14}r_{23}.
$$
This is exactly the same expression obtained for the above expectation value. This 
suggests the following result:
\begin{equation} \label{pfafS}
\langle | \beta_{1} \ldots \beta_{P}  \bar{\beta}_{1} \ldots 
\bar{\beta}_{Q} | \rangle = \textrm{pf} (S_{ij})
\end{equation}
where $S_{ij}$ is the skew symmetric $(P+Q)\times (P+Q)$ matrix 
such that $S_{ij}$ $i<j$ are all the possible contractions
\begin{eqnarray} 
\langle | \beta_{k} \beta_{l}             |\rangle & & \; i,j=1,\ldots,P (k,l=1,\ldots,P) \\
\langle | \beta_{k} \bar{\beta}_{r}       |\rangle & & \; i=1,\ldots,P,j=P+1,\ldots,P+Q (k=1,\ldots,P;\;r=1,\ldots,Q) \\ 
\langle | \bar{\beta}_{r} \bar{\beta}_{s} |\rangle & & \; i,j=P+1,\ldots,P+Q (r,s=1,\ldots,Q) 
\end{eqnarray}
We have also  introduced
another set of fermion operators $\bar{\beta}_i$ that are presumably 
related to the ${\beta}_i$ by some canonical transformation.
The proof of this result can be easily obtained using recursion relations and can also be
easily extended to finite temperature systems \cite{InProg.1}. 

The formula Eq. (\ref{pfafS}) can be readily applied to the problem of computing overlaps between two HFB 
wave functions by noting that  such HFB states can be written as
\begin{equation} 
|w\rangle= {\det C \over \prod_{\alpha=1}^n v_\alpha}\beta_1\beta_2\ldots\beta_{2n} |\rangle  
\label{wquasi}
\end{equation} 
where the normalization factor in front of the product of quasiparticle annihilation operators $\beta_i$ 
contains the occupancies $v_\alpha$ and the determinant of the third transformation in 
the Bloch-Messiah theorem \cite{RS80,BR86} and is constructed to give a normalized $|w\rangle$.
An immediate application of
this result is the formula for the overlap of two HFB states including a 
canonical transformation operator ${\cal R}$ (as the ones that appear when symmetry operations
are applied to the system) acting on one of the states
\begin{equation}
\label{normalo}
\langle w| {\cal R}|w ' \rangle 
=
(-1)^n {\det C^{*} \det C' \over \prod_\alpha^{n} v_\alpha v'_\alpha} {\rm pf}
\left[ \begin{array}{cc}
 V^T          U & V^T        R^T V'^* \\
-V'^\dagger R V & U'^\dagger     V'^*
\end{array}
\right]
\end{equation}
where the matrix $R$ is the representation of the canonical transformation operator ${\mathcal R}$
on the linear Fock space generated by the creation and
annihilation operator $c^\dagger_i$ and $c_i$ in some convenient basis, namely $
{\mathcal R} c^\dagger_i {\mathcal R}^{-1} = \sum_j R_{ij} c^\dagger_j$.

A general multi-quasiparticle overlap including a canonical transformation ${\cal R}$ is easily obtained 
with the previous formalism \cite{PhysRevLett.108.042505}
\begin{eqnarray}
\label{mqp}
&&\langle w| \bar{\beta}_{\mu_r} \cdots \bar{\beta}_{\mu_1} {\cal R} \bar{\beta}'^\dagger_{\nu_1} \cdots \bar{\beta}'^\dagger_{\nu_s} | w'\rangle =
(-1)^n (-1)^{r(r-1)/2} {\det C^{*} \det C' \over \prod_\alpha^{n} v_\alpha^* v'_\alpha} \times \\
&& {\rm pf}
\left[ \begin{array}{cccc}
        V^TU     &         V^T{\bf p}^\dagger          &     V^T R^T {\bf q'}^T        &       V^T R^T V'^*  \\
   -{\bf p}^* V  &     {\bf q}^*  {\bf p}^\dagger      &  {\bf q}^* R^T {\bf q'}^T     &  {\bf q}^* R^T V'^* \\
  -{\bf q}' R V  & -{\bf q}' R {\bf q}^\dagger         &     {\bf p}' {\bf q}'^T       &       {\bf p}' V'^* \\
-V'^\dagger R V  &  - V'^\dagger R{\bf q}^\dagger      &    -V'^\dagger {\bf p}'^T     &  U'^\dagger V'^*    \\
\end{array}\right].
\end{eqnarray}
For this expression to make sense both $r$ and $s$ must have the same number parity. 
The objects ${\bf p}$ and ${\bf q}$ (${\bf p}'$ and ${\bf q}'$) are matrices of 
dimension $r\times 2n$ ($s\times 2n$) with
matrix elements $p_{\mu_j m}  = \bar{V}_{m \mu_j}$ and $q_{\mu_j m} = \bar{U}_{m \mu_j}$. 
This expression has the advantage over the direct application of the
generalized Wick's theorm \cite{Balian.69} that
it avoids the combinatorial explosion of terms to be evaluated.  Namely,
$(r+s-1)!!$ contractions have
to be computed if the multi-quasiparticle overlap is evaluated by
the generalized Wick's theorem.  To give an idea of the 
complexity brought about by the combinatorial explosion, let us just mention, for instance, 
that in the evaluation of the Hamiltonian overlap of two quasiparticle excitations built
on top of an odd-A system, overlaps with ten quasiparticles are required. The number of
terms to be considered if using the GWT would be 9 !! = 945. If two independent two quasiparticle
excitations are considered in each isospin channel the number of quasiparticle operators increases
by four and the number of contractions goes up to a whooping 13 !! = 135 135. 


\section{Gradient method for the HFB equation of odd-A systems}\label{Sec2}

Systems with an odd number of particles  are 
less studied from a theoretical perspective than even-even systems. 
Several circumstances could explain this imbalance and we now discuss
two of them. At the BCS level the wave function of an odd-A system is given by \cite{RS80,BR86}
$$
|\phi_{k_0}\rangle = a^+_{k_0} \prod_{l \neq k_0} (u_l + v_l a^+_l a^+_{\bar{l}} ) | - \rangle
$$
where the orbital labeled $k_0$ is ``blocked". As a consequence, this orbital acquires an occupancy of one and its time reversed 
companion $\bar{k}_0$ becomes empty. Another consequence of blocking, the
fact that the odd-A BCS state is no longer
invariant under time reversal, makes it more difficult to solve the BCS
equations. The Hartree-Fock (HF) and pairing fields also acquire time-odd
components which must be included in the HFB energies and minimization
procedures.

To avoid dealing with the time-reversal breaking issue, 
people have made use of the equal filling
approximation (EFA). It amounts to replace the density matrix and pairing tensor of a blocked 
orbital $k_0$ by a linear combination with equal weights of the density matrices and pairing tensors
of the orbitals $k_0$ and $\bar{k}_0$\footnote{For spherically orbitals the linear combination runs over
the $2j+1$ sub-levels with weights $1/(2j+1)$}. This approximation was widely used even before
the whole procedure was justified as a variational problem on the energy of an statistical
admixture of the $k_0$ and $\bar{k}_0$ blocked states \cite{PhysRevC.78.014304}. Although this procedure
gives results which are very close to the real blocking when the time-odd HF and pairing fields are 
neglected \cite{PhysRevC.81.024316}, the differences with real blocking can amount to a few 
hundred KeV and therefore are relevant for the determination of spin and parities of the ground 
and excited states.

To deal properly with odd-A systems the preferred alternative is the HFB approximation with 
full blocking. The situation 
becomes even more involved than the BCS case because now the odd-A wave function is given by 
$$
|\phi_{\mu_0}\rangle = \alpha^\dagger_{\mu_0} |\phi\rangle
$$
where $\alpha^\dagger_{\mu_0}$ is the quasiparticle creation operator of the quasiparticle 
labeled $\mu_0$ and
$|\phi\rangle$ is the wave function of an even number parity reference system. The reference wave function $|\phi\rangle$
is the vacuum of all the quasiparticle annihilation operators $\alpha_{\mu}$, i.e. $\alpha_{\mu}|\phi\rangle=0$.
On the other hand, $|\phi_{\mu_0}\rangle$ is the vacuum of the set of quasiparticle operators
$$
\alpha_{1},\ldots,\alpha_{\mu_0 -1},\alpha^\dagger_{\mu_0},\alpha_{\mu_0+1},\ldots,\alpha_{N}.
$$
The new quasiparticle vacuum can be obtained from the old one 
\cite{Banerjee1973366,Mang1975325,PhysRevA.79.043602 } by swapping the column $\mu_{0}$ of $U$ and $V$.
This "swapping" procedure is not very easy to incorporate into 
a practical implementation of
the HFB method for odd-A systems.  This is important from a practical
standpoint because of odd-A
systems typically require many HFB calculations with different starting 
wave functions  in order to insure that the ground state is reached \cite{PhysRevC.82.044318,PhysRevC.82.061302}. 
As a consequence, it is very important
to have a robust and efficient method for solving odd-A systems for 
global applications such as the construction of theoretical mass table \cite{PhysRevC.81.024316,Hilaire.05,Afanasjev2011177}.

In the context \cite{InProg.2} of generalizing the approximate
second order gradient method of \cite{PhysRevC.84.014312} it was realized that the ``swapping" 
in the $U$ and $V$ amplitudes can be easily incorporated into the formulas. 
The argument is as follows: the most important object in the HFB method is the generalized 
density matrix 
\begin{equation} 
\mathcal{R}=\left(\begin{array}{cc}
\rho & \kappa\\
-\kappa^{*} & 1-\rho^{*}\end{array}\right)=\left(\begin{array}{cc}
U & V^{*}\\
V & U^{*}\end{array}\right)\left(\begin{array}{cc}
0 & 0\\
0 & \mathbb{I} \end{array}\right)\left(\begin{array}{cc}
U^{+} & V^{+}\\
V^{T} & U^{T}\end{array}\right)=W\mathbb{R}W^{+}\label{eq:R}
\end{equation}
that is given in terms of the unitary Bogoliubov super-matrix
\begin{equation} 
W=\left(\begin{array}{cc}
U & V^{*}\\
V & U^{*}
\end{array}\right)\label{eq:W}
\end{equation}
and the generalized quasi-particle density matrix
\begin{equation} 
\mathbb{R}_{\nu\mu}=\left(\begin{array}{cc}
\left\langle \phi\right|\beta_{\mu}^{\dagger}\beta_{\nu}\left|\phi\right\rangle  & \left\langle \phi\right|\beta_{\mu}\beta_{\nu}\left|\phi\right\rangle \\
\left\langle \phi\right|\beta_{\mu}^{\dagger}\beta_{\nu}^{\dagger}\left|\phi\right\rangle  & \left\langle \phi\right|\beta_{\mu}\beta_{\nu}^{\dagger}\left|\phi\right\rangle \end{array}\right)=
\left(\begin{array}{cc}
0 & 0\\
0 & \mathbb{I}
\end{array}\right).\label{eq:RR}
\end{equation}
When dealing with a blocked HFB state $|\phi_{\mu_0}\rangle$ the 
generalized quasi-particle density matrix becomes
\begin{equation} 
{(\mathbb{R}_{\mu_0})}_{\nu\mu}=\left(\begin{array}{cc}
\left\langle \phi_{\mu_0}\right|\beta_{\mu}^{\dagger}\beta_{\nu}\left|\phi_{\mu_0}\right\rangle  & \left\langle \phi_{\mu_0}\right|\beta_{\mu}\beta_{\nu}\left|\phi_{\mu_0}\right\rangle \\
\left\langle \phi_{\mu_0}\right|\beta_{\mu}^{\dagger}\beta_{\nu}^{\dagger}\left|\phi_{\mu_0}\right\rangle  & \left\langle \phi_{\mu_0}\right|\beta_{\mu}\beta_{\nu}^{\dagger}\left|\phi_{\mu_0}\right\rangle \end{array}\right)=
\left(\begin{array}{cc}
0_{\mu_0} & 0\\
0 & \mathbb{I}_{\mu_0}
\end{array}\right)\label{eq:RRmu0}
\end{equation} 
where the diagonal matrices $0_{\mu_0}$ and $\mathbb{I}_{\mu_0}$ have been introduced. The first of them, $0_{\mu_0}$ 
is zero everywhere except in the position $\mu_0$ of the diagonal. The second is the identity matrix except for
the element $\mu_0$ of the diagonal that is zero. Using now the trivial matrix identity
\begin{equation}\label{eq:TI}
\left(\begin{array}{cc}
0 & 1\\
1 & 0\end{array}\right)
\left(\begin{array}{cc}
1 & 0\\
0 & 0\end{array}\right)
\left(\begin{array}{cc}
0 & 1\\
1 & 0\end{array}\right) = 
\left(\begin{array}{cc}
0 & 0\\
0 & 1\end{array}\right)
\end{equation}
we can write ${(\mathbb{R}_{\mu_0})}$ in terms of $\mathbb{R}$
\begin{equation} 
{(\mathbb{R}_{\mu_0})}=S_{\mu_0} \mathbb{R} S^+_{\mu_0}\label{eq:RRmu0I}
\end{equation} 
by means of a  "swapping" matrix $S_{\mu_0}$ that is inspired by the identity 
of Eq. \ref{eq:TI}. The effect of $S_{\mu_0}$ acting to the left of the
Bogoliubov amplitudes $W$, i.e. ${W}_{\mu_0}=W S_{\mu_0}$, is to swap the
row $\mu_0$ of the $U$ and $V$ amplitudes. The structure of $S_{\mu_0}$ 
is that of an identity matrix except for the rows and columns of the 
label $\mu_0$ in both the $U$ and $V$ blocks. The simplifications implied
by the introduction of $S_{\mu_0}$ can be seen for instance in the 
expression of the generalized density
\begin{equation}\label{eq:Rmu0}
{\mathcal R}_{\mu_0} = W \mathbb{R}_{\mu_0} W^+ = W_{\mu_0} \mathbb{R} W^+_{\mu_0}.
\end{equation}
This tells us that the generalized density can be written in terms of the 
standard formulas (for instance $\rho = V V^T$), but using the new
$U$ and $V$ matrices.
More interesting is the
variation of the energy at first order when the
Bogoliubov amplitudes $W$ are varied according to the most general
canonical transformation (see Ref \cite{PhysRevC.78.014304} for notation)
\begin{equation}
W(\mathbb{Z})= W(0) e^{i\mathbb{Z}}.
\end{equation}
where $\mathbb{Z}$ is an hermitian bipartite matrix 
\begin{equation}
\mathbb{Z}=\left(\begin{array}{cc}
Z^{11} & Z^{20}\\
-Z^{20*} & -Z^{11*}\end{array}\right)\label{eq:Z}.
\end{equation}
The variational parameters of the theory can be enumerated as:
the complex matrix off-diagonal elements $Z_{mn}^{11}$ with $m>n$; the diagonal
$Z_{mm}^{11}$;  and the complex off-diagonal matrix elements $Z_{mn}^{20}$ 
with $m>n$.
The change in energy is given by 
\begin{equation}
\delta E=\frac{i}{2}
\mathrm{Tr_{2}}\left[[\mathbb{R},\mathbb{H}]
\mathbb{Z}\right]+O(\mathbb{Z}^{2})\label{eq:EHFB}
\end{equation}
with 
$$\mathbb{H=}W^{+}(0)\mathcal{H}W(0)=\left(\begin{array}{cc}
H^{11} & H^{20}\\
-H^{20\:*} & -H^{11\:*}\end{array}\right)$$ 
and
$$
\mathcal{H}=\left(\begin{array}{cc}
t+\Gamma & \Delta\\
-\Delta^{*} & -(t+\Gamma)^{*}\end{array}\right).$$
On the other hand, the change in energy for a blocked HFB state $|\phi_{\mu_0}\rangle$ is given
by
\begin{equation}
\delta E_{\mu_0}=\frac{i}{2}
\mathrm{Tr_{2}}\left[[\mathbb{R}_{\mu_0},\mathbb{H}]
\mathbb{Z}\right]+O(\mathbb{Z}^{2})\label{eq:DELTAE}
\end{equation}
where we have replaced $\mathbb{R}$  by $\mathbb{R}_{\mu_0}$
and with an $\mathbb{H}$ computed from the same density.
Using the "swapping" matrix we obtain instead
\begin{equation}
\delta E_{\mu_0}=\frac{i}{2}
\mathrm{Tr_{2}}\left[[\mathbb{R},\mathbb{H}_{\mu_0}]
\mathbb{Z}_{\mu_0}\right]+O(\mathbb{Z}^{2})\label{eq:DELTAEMU0}
\end{equation}
with $\mathbb{H}_{\mu_0}=S_{\mu_0}\mathbb{H}S^+_{\mu_0}=W^{+}_{\mu_0}\mathcal{H}W_{\mu_0}$ and
$\mathbb{Z}_{\mu_0}=S_{\mu_0}\mathbb{Z}S^+_{\mu_0}$.
For the Bogoliubov amplitudes, the following relation is helpful
\begin{equation}\label{eq:WZ}
W(\mathbb{Z})_{\mu_0}= W(0)_{\mu_0} e^{i\mathbb{Z}_{\mu_0}}.
\end{equation}
In practical implementations of the gradient method the exponential in
Eq. \ref{eq:WZ} is computed using the series expansion but corrected
to have unitarity. A possibility is
$$
e^{i\mathbb{Z}_{\mu_0}}\approx  (\mathbb{I} + i\mathbb{Z}_{\mu_0} ) (\mathbb{I} + \mathbb{Z}_{\mu_0} \mathbb{Z}_{\mu_0})^{1/2}.
$$
where the square root of the positive definite matrix is computed by means of
the Cholesky decomposition. Others, based on Pad\'e rational approximations
to the exponential have been explored \cite{PhysRevC.78.014304}.

The previous results are telling us that we can use exactly the same
gradient formalism as in the even-even case but using as 
starting amplitudes $W(0)_{\mu_0}$.
Obviously, the idea can be generalized to multiple quasiparticle excitations
just by adding as many swapping matrices $S_{\mu_0}$ as quasiparticle 
excitations considered.

These ideas are being extended to the expansion of the energy up to
second order required for a "second order" (Newton like) gradient method
and its descendants like the use of the inverse of the sum of quasiparticle
energies $E_\mu+E_\nu$ to damp the "high energy" components of the gradient
$G_{\mu \nu}$ as discussed in\cite{PhysRevC.84.014312} . Although this
little trick can not be used for finite temperature systems (
$\mathbb{R}_{\mu_0}^2=\mathbb{R}_{\mu_0}$ is a necessary condition, not 
satisfied for finite temperature density matrices), work on an efficient
implementation of the gradient method using the inverse of two quasiparticle
energies as a pre-conditioner and valid for any situation (even-A, odd-A or finite temperature)
systems is in progress \cite{InProg.2}.

\section{Conclusions and perspective}

Although the standard BCS theory and its use in nuclear physics are both
more than fifty years old, there are still technical issues, particularly
related to systems with an odd number of particles, that require
further developments to simplify the systematic application of BCS/HFB and
beyond to nuclear systems all over the nuclide chart. In this contribution we have discussed
two of them, one related to the overlaps of HFB wave functions 
required in theories beyond mean field and using the pfaffian of skew-symmetric matrices. The 
other focused on the gradient method with blocked HFB wave functions. 
In the near future, we hope to extend the pfaffian technique to finite
temperature systems and make use of it to simplify the appliction of
of the generalized Wick theorem. Also, approximate second order gradient methods will
be extended to odd-A and finite temperature systems.

\section*{Acknowledgments}
This work was supported by MICINN (Spain) under  
grants Nos. FPA2009-08958, and FIS2009-07277, as well as by  
Consolider-Ingenio 2010 Programs CPAN CSD2007-00042 and MULTIDARK  
CSD2009-00064.  

\bibliographystyle{unsrt}
\bibliography{BCS50}

\end{document}